# Sparse Reconstruction for Radar Imaging based on Quantum Algorithms

Xiaowen Liu, Chen Dong, Ying Luo, Le Kang, Yong Liu, and Qun Zhang

*Abstract*—The sparse-driven radar imaging can obtain the high-resolution images about target scene with the down-sampled data. However, the huge computational complexity of the classical sparse recovery method for the particular situation seriously affects the practicality of the sparse imaging technology. In this paper, this is the first time the quantum algorithms are applied to the image recovery for the radar sparse imaging. Firstly, the radar sparse imaging problem is analyzed and the calculation problem to be solved by quantum algorithms is determined. Then, the corresponding quantum circuit and its parameters are designed to ensure extremely low computational complexity, and the quantum-enhanced reconstruction algorithm for sparse imaging is proposed. Finally, the computational complexity of the proposed method is analyzed, and the simulation experiments with the raw radar data are illustrated to verify the validity of the proposed method.

## I. INTRODUCTION

MICROWAVE imaging via synthetic aperture radar (SAR) and inverse SAR (ISAR) relying on high-resolution performance is playing the pivotal role in military and civilian applications, such as topographic mapping, marine monitoring, and target identification [1]. However, the amount of radar data needs to be collected due to the wide bandwidth signal and long coherent processing interval for high-resolution images and large-scale scene recovery, which result in a grand challenge to data acquisition and storage [2]. For addressing the issue, compressive sensing (CS) theory which is capable of recovering the sparse signals with a high probability from down-sampled data has been widely used in SAR/ISAR imaging with the sparse aperture measurements [3]. Nevertheless, the sparse-driven radar imaging approaches can improve the imaging performance, but most of them need to transmute two-dimensional raw data into a vector, which will lead to significant time-consuming, memory cost, and computational complexity, especially for the situation of high-resolution and large-scale scene imaging [4].

To cope with the critical issue, [1,4,5] has constructed the azimuth-range decouple-based sparse SAR imaging methods which replace the observation matrix in the CS-SAR framework with approximated observations derived from the inverse of traditional matched filtering (MF) based procedures. The sparsity ISAR imaging methods with the block-based CS technique and the Kronecker CS were proposed in [6,7], respectively, to reduce the computational complexity and the memory cost. In addition, a fast sparsity ISAR imaging method was proposed [8], where the enhanced sparsity constraint shrinks the feasible region of the solution so as to reduce the computational cost. Moreover, the ADMM based [9,10] and the smoothed L0 based [11] sparsity recovery methods were proposed, respectively. However, for the sake of low computational complexity and real-time processing, it still needs long-term exertions.

The quantum algorithms profiting from the quantum computer which depends on quantum gates and wires for manipulating the quantum states can bring remarkable exponential speedup or quadratic speedup over some classical algorithms and have been applied in computational chemistry and molecular simulation [12]. Among them the Harrow-Hassidim-Lloyd (HHL) algorithm is proposed in 2009 for solving the linear systems of equations, but there is still no literature about its practical applications and experimental implementations [13]. In order to make the HHL algorithm the useful tool for a specific problem, slightly modifying the quantum circuit design and adding the classical pre/post-processing are necessary to preserve the exponential speedup [14].

In this paper, we are the first to apply the HHL algorithm quantum to the sparse reconstruction problem for radar imaging. After the sparse reconstruction problem about radar imaging is analyzed, the suitable linear equation for the quantum calculation mechanism is presented. Then the quantum circuit and the parameters in the quantum gates are designed, and the quantum-enhanced reconstruction algorithm for sparse imaging is proposed. Finally, the computational complexity is analyzed, and the proposed method and the traditional recovery algorithm are used to acquire the sparse imaging result, respectively. The proposed method can obtain the similar imaging result with that

This work was supported in part by the National Natural Science Foundation of China under Grant 61871396; the Key Research and Development Program of Shaanxi under Grant 2019ZDLGY09-01; the Innovative Talents Promotion Plan in Shaanxi Province under Grant 2020KJXX-011; National University of Defense Technology under Grant 19-QNCXJ. (Corresponding author: Chen Dong and Ying Luo.)

X. W. Liu, C. Dong, and Y. Liu are with the College of Information and Communication, National University of Defense Technology, Xi'an 710100, China. (e-mail: lxw5054@163.com; dongchengfkd@163.com; yongliu@quanta.org.cn). Y. Luo, L. Kang, and Q. Zhang are with the Institute of Information and Navigation, Air Force Engineering University, and with the Collaborative Innovation Center of Information Sensing and Understanding, Xi'an, 710077, China. (e-mail: luoying2002521@163.com; 18810495946@163.com; zhangqunnus@gmail.com).

of the traditional algorithm and has a lower theoretical computational complexity.

## II. SPARSE IMAGING PROBLEM

At first, we briefly introduce the observation-matrix-based sparse reconstruction model for radar imaging. Assuming that full aperture echo data is expressed as $s(t_l, \tau_m)$, $l = 1, 2, \cdots, L_t$, $m = 1, 2, \cdots, M_{all}$, where $t_l$ and $\tau_m$ are fast-time and slow-time sampling sequences, respectively. If the radar only transmits $M_s$ $(M_s < M_{all})$ pulses on the target scene (i.e., $M_s$ effective aperture datum), thus the sparse aperture echo data $s(t_l, \tau_{m'})$ can be expressed as follows:

$$s(t_l, \tau_{m'}) = \sum_{p=1}^{P} \sigma_p \exp\left[ j2\pi f_c (t_l - \frac{2R_p(\tau_{m'})}{c}) \right.$$
$$\left. + j\pi\mu(t_l - \frac{2R_p(\tau_{m'})}{c})^2 \right], t_l \in \left[-\frac{T_p}{2}, \frac{T_p}{2}\right]; m' = 1, 2, \cdots, M_s \quad (1)$$

where parameter $\sigma_p$, $f_c$, $\mu$, and $T_p$ are the scattering coefficient of the p-th scattering point in the scene, the carrier frequency, the chirp rate of radar signal, and the pulse duration, respectively. $R_p(\tau_{m'})$ is the distance between the p-th scattering point and the radar at $\tau_{m'}$. For the two-dimensional matrix $\mathbf{S} \in \mathbb{C}^{L_t \times M_s}$ containing the downsampled echo data $s(t_l, \tau_{m'})$ as elements, its vectorization representation $Y = \{Y(n)\} \in \mathbb{C}^{L_t M_s \times 1}$ can be defined by

$$Y(n) = \sum_{p=1}^{P} \Phi(n, p) \sigma_p, \; n = 1, 2, \ldots, L_t M_s \quad (2)$$

where the elements $\Phi(n, p)$ in the measurement matrix $\mathbf{\Phi} \in \mathbb{C}^{L_t M_s \times P}$ can be given by

$$\Phi(n, p) = \exp\left[ j2\pi f_c (t'_n - \frac{2R_p(\tau'_n)}{c}) + j\pi\mu(t'_n - \frac{2R_p(\tau'_n)}{c})^2 \right] \quad (3)$$

where $t'_n = t_{\lfloor (n-1)/M_s \rfloor + 1}$ and $\tau'_n = \tau_{\text{rem}((n-1)/M_s)+1}$. The operator $\lfloor \cdot \rfloor$ and rem($\cdot$) represent the rounding down operation and the remainder operation, respectively. Let the vector $\boldsymbol{\sigma} \in \mathbb{C}^{P \times 1}$ is the set of the scattering coefficient $\sigma_p$, the downsampled data vector $Y$ can be expressed as follows:

$$Y = \mathbf{\Phi}\boldsymbol{\sigma} + \boldsymbol{n} \quad (4)$$

where the vector $\boldsymbol{n}$ is the noise. Usually, the target scene $\boldsymbol{\sigma}$ is sparse enough that the measurement matrix $\mathbf{\Phi}$ satisfies the condition, such as restricted isometry property (RIP) or mutual incoherence property (MIP). Thus, the target scene $\boldsymbol{\sigma}$ can be exactly reconstructed by solving the sparse imaging problem which is a $L_q$ $(0 \leq q \leq 1)$ optimization problem:

$$\min_{\boldsymbol{\sigma}} \; \|\boldsymbol{\sigma}\|_q \quad \text{s.t.} \quad Y = \mathbf{\Phi}\boldsymbol{\sigma} \quad (5)$$

Nowadays, the sparse recovery algorithms for (5) can be categorized as convex relaxations, non-convex optimization, and greedy algorithms, such as basis pursuit (BP), complex approximate message passing (CAMP), Bayesian compressive sensing (BCS), and orthogonal matching pursuit (OMP).

## III. QUANTUM-ENHANCED RECONSTRUCTION ALGORITHM AND COMPUTATIONAL COMPLEXITY ANALYSIS

### A. Quantum-enhanced Reconstruction Algorithm (QRA)

For the optimization problem as (5), when the target scene sparsity $K$ is roughly estimated, the optimization problem is recognized as the linear least squares problem. Thus, solving the optimization problem (5) is equivalent to calculate the linear equation as $\mathbf{\Phi}^H \mathbf{\Phi} \boldsymbol{\sigma} = \mathbf{\Phi}^H Y$, where $(\cdot)^H$ is the conjugate transpose. However, the linear least squares problem cannot be directly solved by the matrix inversion and multiplication due to the low-rank matrix $\mathbf{\Phi}^H \mathbf{\Phi}$. Thus, the linear equation needs to be transformed, moreover, the new form needs to be eligible for the quantum calculation mechanism to guarantee the high recovery precision. The new linear equation can be formed as

$$\left( \eta \mathbf{\Phi}^H \mathbf{\Phi} + \lambda_0 \mathbf{I} \right) \tilde{\boldsymbol{\sigma}} = \mathbf{\Phi}^H Y \quad (6)$$

where the scaling factor $\eta$ is to resize the eigenvalues of $\mathbf{\Phi}^H \mathbf{\Phi}$ to control the number of qubits and ensure the recovery precision and is also to control the condition number $\kappa$ to ensure a low computational complexity. $\lambda_0$ and I are an arbitrary positive number and the identity matrix, respectively. The sparse processing for $\tilde{\boldsymbol{\sigma}}$ with the appropriate $K$ can keep the error between the final recovery result $\bar{\boldsymbol{\sigma}}$ and $\boldsymbol{\sigma}$ very small. Let $\Xi = \left( \eta \mathbf{\Phi}^H \mathbf{\Phi} + \lambda_0 \mathbf{I} \right)$ and $\boldsymbol{\gamma} = \mathbf{\Phi}^H Y$, and thus (6) can be rewritten as $\Xi \tilde{\boldsymbol{\sigma}} = \boldsymbol{\gamma}$. According to (3) and (6), we can draw the conclusion that the matrix $\Xi$ is Hermitian and has $s$ nonzero entries per row.

As mentioned above, the quantum algorithms can achieve exponential speedup or quadratic speedup over some classical algorithms. Among them, the HHL algorithm is a kind of the quantum algorithm for linear systems of equations usually expressed as $\mathbf{A}\boldsymbol{x} = \boldsymbol{b}$ and requires that the matrix $\mathbf{A}$ is Hermitian, $s$-sparse, and efficiently row computable. Considering the characteristic of the matrix $\Xi$, after normalizing the vector $\boldsymbol{\gamma}$, the target scene can be obtained by solving the linear equation $\Xi \hat{\boldsymbol{\sigma}} = \hat{\boldsymbol{\gamma}}$ using the HHL algorithm, where $\hat{\boldsymbol{\gamma}}$ and $\hat{\boldsymbol{\sigma}}$ are the normalized vector of $\boldsymbol{\gamma}$ and the product of the vector $\tilde{\boldsymbol{\sigma}}$ and the normalized coefficient of $\boldsymbol{\gamma}$, respectively. The quantum-enhanced reconstruction algorithm for radar imaging can be summarized as Algorithm 1.

---

**Algorithm 1** Quantum-enhanced reconstruction algorithm

**Input:** the measurement matrix $\Phi$ and the downsampled data vector $Y$.
1. Select the parameters $\lambda_0$ and $\eta$ to construction the matrix $\Xi$.
2. Design the quantum circuit for $\Xi\hat{\sigma} = \hat{\gamma}$ according to $\Xi$ and $\hat{\gamma}$, and set the parameters of the quantum gates in the circuit.
3. Prepare the initial state $|b\rangle = \sum_{i=0}^{N_I - 1} \hat{\gamma}_{i+1} |i\rangle$ in the Input register $I$.
4. Perform the designed quantum circuit to obtain the outcome $|x\rangle = \hat{\sigma}$.
5. Extract the $K$ largest elements in $\tilde{\sigma}$ to form the recovery result $\bar{\sigma}$.

---

The quantum circuit and parameter selection for the sparse imaging problem is designed as illustrated in Fig. 1. The quantum circuit contains five different kinds of registers, i.e., Ancilla register $S$, register $A$, register $B$, register $C$, and Input register $I$, and can be separated into the three stages: phase estimation, controlled rotation, and uncomputation. Quite apart from register $I$, the initial quantum state of other registers is $|0\rangle^{\otimes n_r}$, and $n_r$ which represents the number of qubits in the registers may be different for the different registers. The initial state of register $I$ need to be prepared as the unit vector $\hat{\gamma}$, i.e., the initial state $|b\rangle = \sum_{i=0}^{N_I - 1} \hat{\gamma}_{i+1} |i\rangle$, where $\hat{\gamma}_{i+1}$ is the (i+1)-th element in $\hat{\gamma}$, $|i\rangle$ is the basis state of register $I$, and $N_I$ is the dimension of $\hat{\gamma}$. Thus, the number of qubits in the register $I$ is $\lceil \log_2 N_I \rceil$, where $\lceil \cdot \rceil$ is the rounding up operation, so that the theoretical computational complexity of the recovery algorithm for the sparse imaging problem may potentially and obviously be reduced, especially when the dimension of $\hat{\gamma}$ is extremely large.

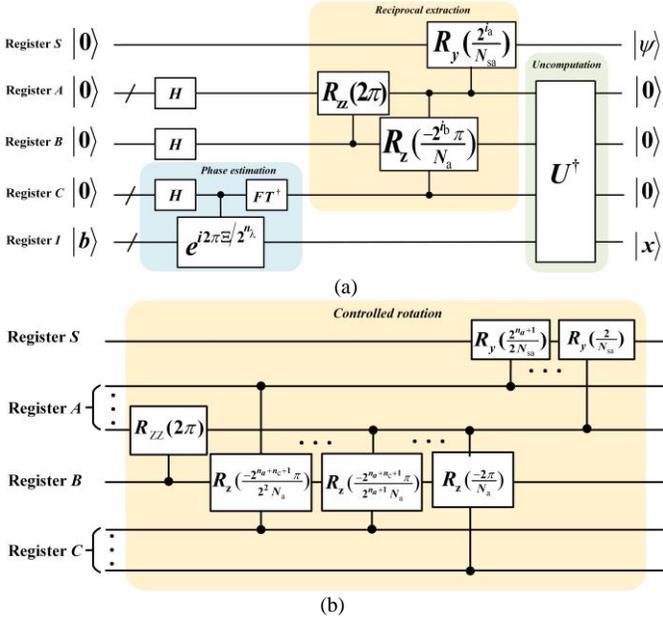

Fig. 1. Quantum circuit of the HHL algorithm for the sparse imaging problem. (a) is the overall circuit design. (b) is the detail of the controlled rotation part.

In the stage of well-known quantum phase estimation, the first quantum gate, i.e., Hadamard gate, takes the state $|0\rangle^{\otimes n_c}$ in the register $C$ to the superposition state $\sum_{i=0}^{2^{n_c}-1} |i\rangle / \sqrt{2^{n_c}}$, where $n_c$ effecting the precision of the eigenvalues of $\Xi$ is the number of qubits in the register $C$. Applying the unitary operator $e^{i2\pi \Xi / 2^{n_\lambda}}$ to the state $|b\rangle$ is implemented using the second quantum operation achieved by Hamiltonian simulation. $n_\lambda$ is the bit number of the binary integer value regarding the maximum eigenvalue of $\Xi$. After the former two steps, the initial state is transformed into $\sum_{j=1}^{n_j} \sum_{k=0}^{2^{n_c}-1} \beta_j e^{i 2\pi \lambda_j k / 2^{n_\lambda}} |k\rangle^C |u_j\rangle^I / \sqrt{2^{n_c}}$, where the superscript in the state, such as $|\cdot\rangle^C$, indicates the register storing the state. $\lambda_j$ represents an eigenvalue of $\Xi$, and $n_j$ is the number of the eigenvalues. Meanwhile, the initial state $|b\rangle^I$ is decomposed in the eigenvector of $\Xi$ so that the state $|b\rangle^I$ becomes $\sum_{j=1}^{n_j} \beta_j |u_j\rangle^I$, where $|b\rangle^I = \sum_{j=1}^{n_j} \beta_j |u_j\rangle^I$ and the vector representation of the state $|u_j\rangle^I$ is the corresponding eigenvector of $\Xi$. The third quantum operation in the stage is inverse quantum Fourier transform achieved by a series of controlled-$R_z$ gates can move the phase information $\lambda_j$ from the probability amplitude to the quantum bases. Therefore, the quantum state in the register $C$ and register $I$ after quantum phase estimation implements the evolution as follows:

$$|0\rangle^{\otimes n_r} |b\rangle \mapsto \sum_{j=1}^{n_j} \beta_j |\tilde{\lambda}_j\rangle^C |u_j\rangle^I \qquad (7)$$

where $\tilde{\lambda}_j = 2^{(n_c - n_\lambda)} \lambda_j$ in binary format.

In the stage of the controlled rotation, the first quantum operation implemented by a series of controlled-$R_{zz}$ gates maps the state $\sum_{p=0}^{1} \sum_{l=0}^{2^{n_a}-1} |l\rangle^A |p\rangle^B / \sqrt{2^{n_a+1}}$ in the register $A$ and $B$ to $\sum_{p=0}^{1} \sum_{l=0}^{2^{n_a}-1} e^{i 2\pi p} |l\rangle^A |p\rangle^B / \sqrt{2^{n_a+1}}$. $n_a$ is the number of qubits in the register $A$ and its size is related to the lowest common multiple $N_a$ of the scaling eigenvalues $\{\tilde{\lambda}_j\}$, while the number of qubits in the register $B$ is 1. The second quantum operation is to apply the controlled-$R_z$ gates to the state in the register $B$ controlled by the qubits in the register $A$ and $C$. After this step, for a certain eigenvalue, the state in the register $A$, $B$, and $C$ becomes as

$$\sum_{p=0}^{1} \sum_{l=0}^{2^{n_a}-1} e^{i \frac{2\pi p}{N_a}(N_a - l\tilde{\lambda}_j)} |l\rangle^A |p\rangle^B |\tilde{\lambda}_j\rangle^C / \sqrt{2^{n_a+1}} \qquad (8)$$

with the constraint $N_a - l\tilde{\lambda}_j = 0$ to make the state vanish, thus the value of state $|l\rangle^A$ is $l = N_a / \tilde{\lambda}_j$. Finally, a series of $R_y$ gates controlled by the state $|l\rangle^A$ are acted on the ancilla register $S$ so that the ancilla state becomes as $\cos(l/N_{sa})|0\rangle+\sin(l/N_{sa})|1\rangle$. Due to the arbitrary large integer $N_{sa}$, the state can be written as

$$\sum_{j=1}^{n_j}\sum_{p=0}^{1}\frac{\beta_j}{\sqrt{2^{n_a+1}}}\left(\sqrt{1-\left(\frac{N_a}{\tilde{\lambda}_j N_{sa}}\right)^2}|0\rangle+\frac{N_a}{\tilde{\lambda}_j N_{sa}}|1\rangle\right)\left|\frac{N_a}{\tilde{\lambda}_j}\right\rangle|p\rangle|\tilde{\lambda}_j\rangle|u_j\rangle \quad (9)$$

After the final stage, i.e., uncomputation, the state in the register $A$, $B$, and $C$ is set back to the initial state $|0\rangle^{\otimes n_r}$, thus the final state of the quantum circuit takes the form as

$$\sum_{j=1}^{n_j}\beta_j\left(\sqrt{1-\left(\frac{N_a}{\tilde{\lambda}_j N_{sa}}\right)^2}|0\rangle+\frac{N_a}{\tilde{\lambda}_j N_{sa}}|1\rangle\right)|0\rangle^A|0\rangle^B|0\rangle^C|u_j\rangle \quad (10)$$

For (10), when we measure the ancilla state and obtain $|1\rangle$, the state in the register $I$ is $|x\rangle=\sum_{j=1}^{n_j}C\beta_j|u_j\rangle/\tilde{\lambda}_j$ which is proportional to the solution of $\Xi\tilde{\sigma}=\gamma$.

*B. Computational Complexity Analysis*

When the range profile is obtained and the range cell migration is corrected, the measurement matrix $\Phi$ is designed based on the product of sparse sampling matrix and inverse DFT matrix, and the high-resolution image is reconstructed by sparse recovery in each range cell. Suppose the dimensionality of $\Phi$ is $M_s\times M_{all}$ and the target scene sparsity in each range cell is $K_c$, the computational complexity of the OMP algorithm used to reconstruct the target image is $O(K_c L_t M_{all} M_s)$, while the computational complexity of the QRA for the sparse imaging problem is $O(\kappa L_t \log(M_{all})/\epsilon)$. The factor $\kappa$ is the condition number of $\Xi$, and $\epsilon$ is the reconstruction error in the output state $|x\rangle$. It is obviously that the computational complexity brought by the data in the cross-range direction is reduced exponentially, thus the quantum-enhanced method is beneficial to the imaging problem with the long coherent processing interval. If the image is reconstructed by processing the data in the whole range cell, the computational complexity of the quantum method is $O(\kappa\log(L_t^2 M_{all} M_s)/\epsilon)$. When the appropriate matrix $\Xi$ and number of qubits are selected to guarantee $\kappa$ and $\epsilon$ to be $\mathrm{poly}\log(M_{all})$ or $\mathrm{poly}\log(L_t^2 M_{all} M_s)$, the QRA method can roughly achieve an exponential speedup.

IV. EXPERIMENTAL RESULTS AND ANALYSIS

In this section, the raw radar data of the F-16 model with 1:8 scaling factor in microwave anechoic chamber and the Yak-42 airplane are processed, respectively, by the proposed method to verify its reconstruction performance.

In the all-metal F-16 scaling model experiment, the central azimuth, synthetic aperture angle, and sampling interval are $180°$, $5°$, and $0.04°$, respectively. The frequency range and the number of frequency sampling points are selected as 34.2857~37.9428 GHz and 401, respectively. To obtain the ideal complexity and precision, the factors $\eta$ and $\lambda_0$ are set as 23 and 1 to bring the appropriate values about the eigenvalues $\{\tilde{\lambda}_j\}$ of $\Xi$ for the required qubit number. In the designed quantum circuit, the register $A$, $C$, and $I$ contain 2, 3, and 7 qubits, respectively. The sparse imaging results (as shown in Fig. 2) of the OMP algorithm and the QRA are calculated, and their lowest computational complexity are $O(10^7)$ and $O(10^2)$, respectively.

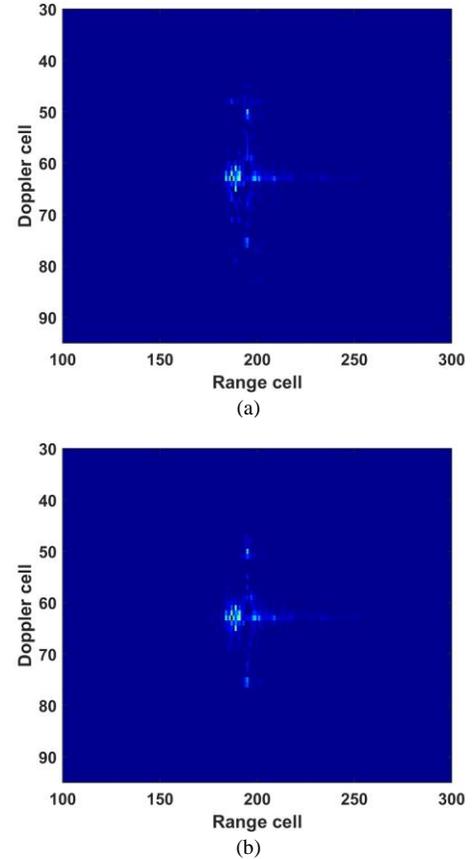

Fig. 2. The sparse reconstruction results regarding the F-16 model. (a) reconstructed by the QRA. (b) reconstructed by the OMP algorithm.

In the Yak-42 experiment, the center frequency, bandwidth, and PRF are 5.52 GHz, 400 MHz, and 400 Hz. The factors $\eta$ and $\lambda_0$ are set as 33 and 1 and the register $A$, $C$, and $I$ contain 2, 3, and 8 qubits, respectively. The sparse imaging results (as shown in Fig. 3) are acquired by the classical and quantum-enhanced method, and their lowest computational complexity are $O(10^8)$ and $O(10^2)$, respectively. From Fig. 2 and Fig. 3, the sparse imaging results obtained by the quantum-enhanced method are very similar with that of the classical method.

Finally, the root mean square error (RMSE) between the reconstructed image $\bar{\sigma}$ and the original scattering coefficient $\sigma$, defined as $\|\bar{\sigma}-\sigma\|_2 / \|\sigma\|_2$, to quantitate the reconstruction error, and the RMSE of the classical and quantum-enhanced reconstruction results is illustrated in Fig. 4. The RMSE of the quantum-enhanced recovery algorithm is slightly higher than that of the classical recovery algorithm, but it is might be acceptable and it is of great potential to further improve the recovery precision of the quantum-enhance method.

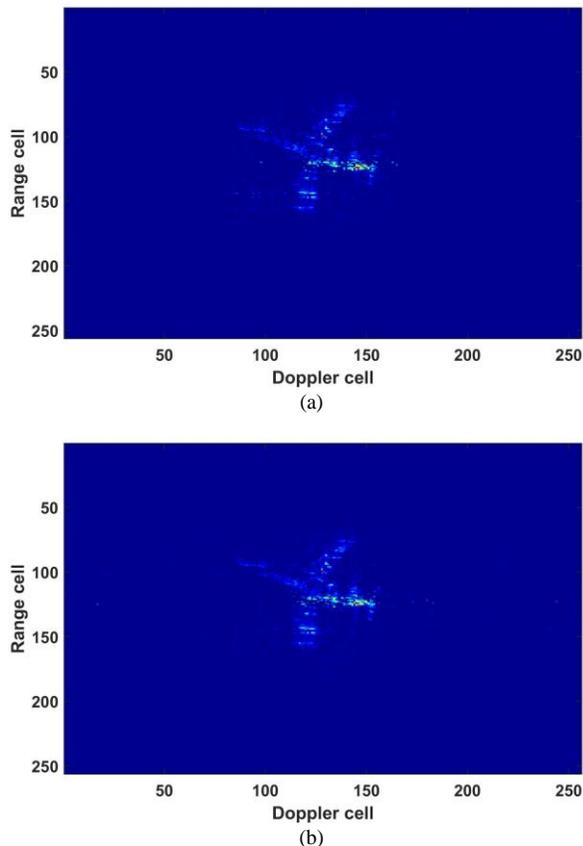

Fig. 3. The sparse reconstruction results regarding the Yak-42 airplane. (a) reconstructed by the QRA. (b) reconstructed by the OMP algorithm.

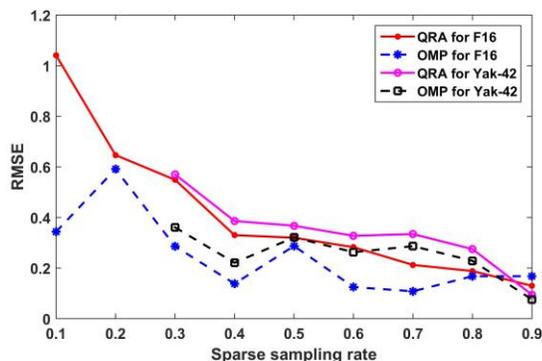

Fig. 4. The RMSE comparison between the classical and quantum-enhanced method.

## V. CONCLUSION

In this paper, a quantum-enhanced sparse reconstruction method for radar imaging is proposed. After analyzing the sparse imaging problem and determining the mathematical problem to be solved, the algorithm flow of QRA and the quantum circuit with the appropriate parameters is presented to ensure a low system complexity. The radar data collected from microwave anechoic chamber and the real airplane echo data are processed to verify the performance of the proposed method. Comparing with the classical algorithm, the similar reconstructed images are acquired by the proposed method, and its construction error quantified by the RMSE did not differ much in high sparsity sampling rate. By theoretically analyzing on the computational complexity, the conclusion that the quantum-enhanced sparse reconstruction method achieves an approximate exponential speedup is verified.

REFERENCES

[1] H. Bi, G. A. Bi, B. C. Zhang, and W. Hong, "Complex-Image-Based Sparse SAR Imaging and Its Equivalence," *IEEE Trans. Geosci. Remote Sens.*, vol. 56, no. 9, pp. 5006–5014, Sep. 2018.

[2] G. Xu, L. Yang, G. A. Bi, and M. D. Xing, "Enhanced ISAR Imaging and Motion Estimation With Parametric and Dynamic Sparse Bayesian Learning," *IEEE Trans. Comput. Imag.*, vol. 3, no. 4, pp. 940-952, Dec. 2017.

[3] M. Cetin *et al.*, "Sparsity-driven synthetic aperture radar imaging: Reconstruction, autofocusing, moving targets, and compressed sensing," *IEEE Signal Process. Mag.*, vol. 31, no. 4, pp. 27–40, Jul. 2014.

[4] H. Bi, G. A. Bi, B. C. Zhang, W. Hong, and Y. R. Wu, "From Theory to Application: Real-Time Sparse SAR Imaging," *IEEE Trans. Geosci. Remote Sens.*, vol. 58, no. 4, pp. 2928–2936, Apr. 2020.

[5] J. Fang, Z. B. Xu, B. C. Zhang, W. Hong, and Y. R. Wu, "Fast Compressed Sensing SAR Imaging Based on Approximated Observation," *IEEE J. Sel. Topics Appl. Earth Observ. Remote Sens.*, vol. 7, no. 1, pp. 354–363, Jan. 2014.

[6] J. Fang, L. Z. Zhang, and H. B. Li, "Two-Dimensional Pattern-Coupled Sparse Bayesian Learning via Generalized Approximate Message Passing," *IEEE Trans. Image Process.*, vol. 25, no. 6, pp. 2920-2930, Jun. 2016.

[7] W. Qiu, J. X. Zhou, and Q. Fu, "Tensor Representation for Three-Dimensional Radar Target Imaging With Sparsely Sampled Data," *IEEE Trans. Comput. Imag.*, vol. 6, pp. 263-275, Jan. 2020.

[8] G. H. Zhao, F. F. Shen, J. Lin, G. M. Shi, and Y. Niu, "Fast ISAR Imaging Based on Enhanced Sparse Representation Model," *IEEE Trans. Antennas Propag.*, vol. 65, no. 10, pp. 5453-5461, Oct. 2017.

[9] S. H. Zhang, Y. X. Liu, and X. Li, "Fast Sparse Aperture ISAR Autofocusing and Imaging via ADMM Based Sparse Bayesian Learning," *IEEE Trans. Image Process.*, vol. 29, pp. 3213-3226, Jun. 2020.

[10] C. Y. Hu, Z. Li, L. Wang, J. Guo, and O. Loffeld, "Inverse Synthetic Aperture Radar Imaging Using a Deep ADMM Network," in *Proc. IRS*, Ulm, Germany, 2019, pp. 1-9.

[11] E. Giusti, D. Cataldo, A. Bacci, S. Tomei, and M. Martorella, "ISAR image resolution enhancement: Compressive sensing versus state-of-theart super-resolution techniques," *IEEE Trans. Aerosp. Electron. Syst.*, vol. 54, no. 4, pp. 1983–1997, Aug. 2018.

[12] M. A. Nielsen and I. L. Chuang, "Quantum Computation and Quantum Information: 10th anniversary edition," Cambridge University Press and Tsinghua University Press, 2015.

[13] A. W. Harrow, A. Hassidim, and S. Lloyd, "Quantum Algorithm for Linear Systems of Equations," *Phys. Rev. Lett.*, vol. 103, pp. 150502, Oct. 2009.

[14] S. Aaronson, "Read the fine print," *Nat. Phys.*, vol. 11, no. 4, pp. 291-293, Apr. 2015.